\documentclass[twocolumn,preprintnumbers,amssymb,amsmath,superscriptaddress,letterpaper]{revtex4}[10pt]

\usepackage{graphicx}
\usepackage{dcolumn}
\usepackage{bm}
\usepackage{natbib}
\usepackage{pstricks}

\newcommand{\be}{\begin{equation}}
\newcommand{\ee}{\end{equation}}
\newcommand{\ba}{\begin{eqnarray}}
\newcommand{\ea}{\end{eqnarray}}
\newcommand{\nn}{\nonumber}

\newcommand{\cm}{{\rm ~cm}}

\newcommand{\MHz}{{\rm ~MHz}}

\newcommand{\kpc}{{\rm ~kpc}}

\begin{document}

\title{Three Dimensional Distribution of Atomic Hydrogen in the Milky Way }
\author{Maryam Tavakoli}\email{tavakoli@sissa.it}
\affiliation{SISSA, Via Bonomea, 265, 34136 Trieste, Italy}
\affiliation{INFN, Sezione di Trieste, Via Bonomea 265, 34136 Trieste, Italy}

\date{\today}

\begin{abstract}
A new model for three dimensional distribution of atomic hydrogen gas in the Milky Way is derived using the 21cm LAB survey data. The global features of the gas distribution such as spiral arms are reproduced. The Galactic plane warps outside the solar orbit and the thickness of the gas disk flares outward the Galaxy. It is found that the mass of atomic hydrogen gas within a radius of 20 kpc is $4.3 \times 10^9 M_{\odot}$.
\end{abstract}

\maketitle


\section{Introduction}

The distribution of gas in the Milky Way reveals the global structure and dynamics of the interstellar medium \cite{2003ApJ...588..805K, 2009ARA&A..47...27K}. Furthermore, the fragmentation of cosmic rays during their propagation within Galaxy is caused by interactions with interstellar gas. Besides, the diffuse $\gamma$-ray components produced by decays of neutral pions and bremsstrahlung of electrons and positrons are correlated to the gas distribution in the Galaxy.  One of the major constituents of the interstellar gas is atomic hydrogen which is traced by its 21cm line emission. Several models for distribution of atomic hydrogen in our Galaxy have been constructed \cite{Nakanishi:2003eb, 1990ARA&A..28..215D, 1976ApJ...208..346G}. However, recent all sky 21cm surveys with high angular resolution, strong sensitivity and large velocity range motivate us to devise a new model with detailed features.

In this note, we construct a model for three dimensional distribution of atomic hydrogen gas in our Galaxy. To that end, the Leiden-Argentine-Bonn (LAB) survey data \cite{Kalberla:2005ts} is used. This survey merges the Instituto Argentino de Radioastronomia (IAR) southern sky survey \cite{2000A&AS..142...35A, Bajaja:2005tn} with the Leiden/Dwingeloo Survey (LDS) \cite{Hartmann:1997}. The data are corrected for the stray radiation. This survey has an angular resolution of $0.6^{\circ}$ and a velocity sampling of 1 km/s, with velocity range of (-450,400) km/s. It is presently the most sensitive 21cm line survey with the most extensive spatial and kinematic coverage.

The derived distribution of atomic hydrogen gas can be applied for estimating the interaction rate of cosmic rays during their propagation as well as evaluating the diffuse gamma rays emission. Small scale features of the gas distribution manifest themselves in diffuse gamma ray sky maps. These structures can be traced in high angular resolution maps of the {\it Fermi} gamma ray telescope.   
  
This paper is organized as follows; In section \ref{sec.21cm_line} the properties of 21cm line and its connection to the atomic hydrogen number density is reviewed. In section \ref{sec.method} the method of deriving the gas density in the outer and inner part of the solar orbit and at tangent points is explained.  We also discuss the assumptions on the rotation curves.  In section \ref{sec.results} we present the general properties of the gas distribution and finally, we conclude in section \ref{sec.summary}.


\section{The $21\cm$ Line Emission}
\label{sec.21cm_line}

Hydrogen atom emits line radiation with wavelength $\lambda_0=21.1\cm~(\nu_0=1420.4058\MHz)$ through a hyperfine transition when the spins of electron and proton flip from being parallel to antiparallel. Since the transition probability is too small, collisions have enough time to establish an equilibrium distribution of hydrogen atoms in the upper and lower states labelled 2 and 1 respectively.  Thus applying the Boltzmann distribution the ratio of the number of atoms in these states is given by 
\be
\frac{n_2}{n_1}=\frac{g_2}{g_1}\exp{(-h\nu_0/kT_s)} \approx \frac{g_2}{g_1}(1-\frac{h\nu_0}{kT_s}), \label{eq.Boltzmann}
\ee
where $g_2$ and $g_1$ are, respectively, the statistical weights of the upper and lower levels with ratio of $g_2/g_1=3$. The excitation temperature $T_s$  is called the spin temperature and under most circumstances $T_s \gg h\nu_0/k=7\times10^{-2}~K$.  
The radiation transfer equation in terms of the radiation intensity $I_{\nu}$, namely the radiant energy per second per unit area per steradian per bandwidth, can be expressed by (for more details see \cite{Longair, Binney, Mihalas1981,Choudhuri})
\be
\frac{dI_{\nu}}{dx}=\frac{k_{\nu}}{4\pi}-\chi_{\nu}I_{\nu}.  
\ee
The increase in intensity, in traversing $dx$,  is $k_{\nu}/4\pi$ where $k_{\nu}$ is the emissivity of the plasma. The decrease in intensity in the same distance increment is $\chi_{\nu}I_{\nu}$ where $\chi_{\nu}$ is the absorption per unit path length. 

The emissivity and absorption per unit frequency interval in terms of the line width of the neutral hydrogen profile $\delta \nu$ are
\ba
k_{\nu} &=&n_2(x)A_{21}h\nu_0/\delta \nu, \\
\chi_{\nu}&=&\frac{1}{4\pi}(n_1B_{12} - n_2B_{21})h\nu_0/\delta \nu. 
\label{eq.absorption}\ea
The Einstein's coefficients $A_{21}$, $B_{12}$ and $B_{21}$ are the intrinsic properties of atoms and satisfy the following relations
\be
\frac{B_{21}}{B_{12}}=\frac{g_1}{g_2}, \quad \frac{A_{21}}{B_{21}}=\frac{2h\nu_0^3}{c^2}.
\label{eq.Einstein_Coefficient} \ee
Using \ref{eq.Boltzmann} and \ref{eq.Einstein_Coefficient}, the absorption (\ref{eq.absorption}) can be rewritten as follows 
\be
\chi_{\nu}\delta\nu=\frac{3hc^2}{32\pi kT_s\nu_0}A_{21}n_H, 
\label{eq.absorption2}\ee
where we have used the fact that the number density of hydrogen atoms in the lower state is $1/4$ of the total hydrogen number density. 
The  optical depth in terms of absorption $\chi_{\nu}$ is
\be
\tau_{\nu}=\chi_{\nu}\delta r,
\ee
where $\delta r$ is the path length. The optical depth, itself, is used to define the brightness temperature   
\be
T_b \equiv T_s(1-e^{-\tau_{\nu}}).
\ee
The line width $\delta \nu$ is caused by the motions of hydrogen gas with respect to the observer. Thus it can be replaced by $\frac{\nu_0}{c}\delta v_r$ where $v_r$ is the gas radial velocity relative to the Sun. Assuming that the motion of gas around the Galactic center is purely circular, the gas located at Galactocentric radius $R$ and altitude $|z|$ has a radial velocity with respect to the Sun which is given by
\be
v_r=\Big(\frac{\theta(R,z)}{R} - \frac{\theta_{\odot}}{R_{\odot}}\Big)R_{\odot}\sin l \cos b.
\label{eq.radial_velocity}\ee
In the above, $\theta(R,z)$ is the rotation velocity of the gas, $l$ and $b$ are, respectively, the longitude and latitude. Eq. (\ref{eq.absorption2}) is then written as
\be
n_H\delta r= CT_s\ln{(1-\frac{T_b}{T_s})}\delta v_r
\label{eq.volume_density}\ee
where the constant $C$ is
\be
\begin{split} C=&-\frac{32\pi k\nu_0^2}{3hc^3A_{21}}\cr =&-1.823\times 10^{18} cm^{-2}K^{-1}(km/s)^{-1}.
\end{split}\ee
The column density is obtained by integrating eq.(\ref{eq.volume_density}) over path lengths and radial velocities along a line of sight. 
\be
N_H[cm^{-2}]=\int_{l.o.s}\!\!\!n_Hdr=\int_{l.o.s}\!\!\!CT_s\ln{\Big(1-\frac{T_b}{T_s}\Big)}dv_r. 
\ee
The measurement of the brightness temperature $T_b$ over a large range of radial velocities and directions in the sky is provided by the combined LAB survey \cite{Kalberla:2005ts}. 

The spin temperature is much greater than the brightness temperature for optically thin ($\tau_{\nu} \ll 1$) 21cm line emission. However, toward extragalactic sources $T_s$ varies from 40 to 300 K, depending on the location and velocity \cite{Strasser:2004uh}.  Moreover, outside the solar circle up to Galactocentric radius of 25 kpc the spin temperature is in the range of 250 to 400 K \cite{Dickey:2009mu}. 
We assume a globally constant spin temperature equal to 150 K, which is also the maximum observed $T_b$ in the LAB survey data. Although using different values of $T_s$ may change the column density of atomic hydrogen \cite{Johannesson:2010fr}, the whole structure of the gas remains unchanged. 


\section{Derivation of Atomic Hydrogen Number Density}
\label{sec.method}

The atomic hydrogen number density at a given heliocentric distance $r$, longitude $l$ and latitude $b$ is derived from (\ref{eq.volume_density}) as
\be
n_H(r,l,b)[cm^{-3}]= CT_s\ln{\Big(1-\frac{T_b(l,b,v_r)}{T_s}\Big)}\Big|\frac{\delta v_r}{\delta r}\Big|
\label{eq.volume_density2}\ee
where $l=0,~b=0$ corresponds to the Galactic center.

The rotation velocity of the gas $\theta(R,z)$ away from the Galactic plane becomes smaller than that of the gas at the underlying disk. At small Galactocentric radii the altitude dependence of rotation curves  is prominent while it becomes less important at large values of $R$ \cite{Kalberla:2007sr}. Moreover, the vertical extension of hydrogen gas is small inside the solar circle and it increases outward the Galaxy. For these reasons, it is a valid assumption to ignore the lagging rotation.

Inside the solar circle ($R<R_{\odot}$) we use the rotation curve of \cite{Clemens:1985} which is fitted by a polynomial of the form
\be
\theta(R)=\Sigma_{n=0}^7A_nR^n. 
\ee
The coefficients $A_n$ are obtained by assuming $R_{\odot}=8.5\kpc$ and $\theta_{\odot} = 220$ km/s. For $R>R_{\odot}$ there is a general consensus that it is a fair approximation to assume a flat rotation curve with $\theta = \theta_{\odot}$ \cite{Levine:2006ty, McClureGriffiths:2007ts}. It is worth noting that the angular velocity
\be 
\omega(R)=\frac{\theta(R)}{R}=\frac{v_r}{R_{\odot}\sin l \cos b}+\omega_{\odot}, 
\label{eq.angular_velocity}\ee 
is always positive and increases toward the Galactic center.  Radial velocities giving negative $\omega$ are forbidden. They correspond to the peculiar motions of the local gas.

The derivative of radial velocity with respect to heliocentric distance $\frac{\delta v_r}{\delta r}$ is computed by using the chain rule as follows
\ba
\frac{\delta v_r}{\delta r}&=& \frac{\delta v_r}{\delta \omega}\frac{\delta \omega}{\delta R}\frac{\delta R}{\delta r}  \label{eq.jacobian} \\
                                         & =&R_ {\odot}\sin l \cos^2 b(r \cos b - R_{\odot} \cos l)\frac{1}{R}\frac{\delta \omega}{\delta R}. 
\nn\ea
Therefore, the gas number density at every given $(r,l,b)$ is obtained by inserting  eq.(\ref{eq.jacobian}) in eq.(\ref{eq.volume_density2}). The heliocentric distance $r$ associated to $(l,b,v_r)$ or $(l,b,R)$ is determined by                                                                                                
\begin{subequations}
\be r=\frac{R_{\odot}\cos l \pm \sqrt{R^2-R_{\odot}^2\sin ^2l}}{\cos b}  \quad  R \le R_{\odot} \ee
\be r=\frac{R_{\odot}\cos l + \sqrt{R^2-R_{\odot}^2\sin ^2l}}{\cos b}  \quad  R>R_{\odot}. \ee
\label{eq:heliocentric_distance}\end{subequations}
In the inner part of the solar orbit there are two kinematically allowed distances (except for the tangent points $R=R_{\odot}|\sin l|$ where they coincide) and for the outer part there is only one. Although there is no distance ambiguity at tangent points, because $\frac{\delta v_r}{\delta r}$ is zero eq.(\ref{eq.volume_density2}) fails to determine the gas density. We describe the method of obtaining the gas density at tangent points in section \ref{sec.tangent_points}. In the case of distance degeneracy the observed intensity must be distributed among the near-far points as explained in section \ref{sec.inner_galaxy}. Note that $\frac{\delta v_r}{\delta r}$ is also zero toward the Galactic center/anti-center and right above/below the Sun. The number density of points in these directions is calculated by linear interpolation between $n_H(r,l,b)$ of  nearby points. 


\subsection{Tangent Points}
\label{sec.tangent_points}

The closest point to the Galactic center at every given direction $r\cos b = R_{\odot}\cos l$, has extreme positive (in the first quadrant) or negative (in the fourth quadrant) radial velocity which is called terminal velocity $v_t$. Due to velocity dispersion $\sigma_v$, the velocity profiles do not have a sharp cutoff at tangent points. The emission from the tangent points is ideally a bivariate Gaussian in altitude and velocity. But emission from the nearby radii are not well separated in velocity because of the velocity dispersion. Atomic hydrogen gas in the vicinity of the tangent point has radial velocity with $|v_r| \ge |v_t|-\sigma_v$. The number density around the tangent point is obtained by dividing the emission from this velocity range ($|v_r| \ge |v_t|-\sigma_v$)  by the corresponding path length  \cite{Nakanishi:2003eb}. In the first quadrant, it is
\begin{subequations}
\be n_H(r,l,b) = \frac{CT_s\int_{v_t - \sigma_v}^{\infty}\ln\Big(1-\frac{T_b(l,b,v_r)}{T_s}\Big) dv}{r_2(v_t - \sigma_v) - r_1(v_t - \sigma_v)},  \label{eq.first}\ee
and in the fourth quadrant it is 
\be n_H(r,l,b) = \frac{CT_s\int_{- \infty}^{v_t + \sigma_v}\ln\Big(1-\frac{T_b(l,b,v_r)}{T_s}\Big) dv}{r_2(v_t + \sigma_v) - r_1(v_t + \sigma_v)}. \label{eq.fourth}
\ee\end{subequations}
In the above, $r_1$ and $r_2$ are, respectively, the near and far heliocentric distances associated to radial velocity of $v_t-\sigma_v (v_t+\sigma_v$) in the first (fourth) quadrant. The velocity dispersion has been estimated to be about 9 km/s for the first quadrant and 9.2 km/s for the fourth quadrant \cite{Malhotra:1994qj}, however it is larger close to the Galactic center.   

\begin{figure}
\begin{center}
\includegraphics[scale=.5]{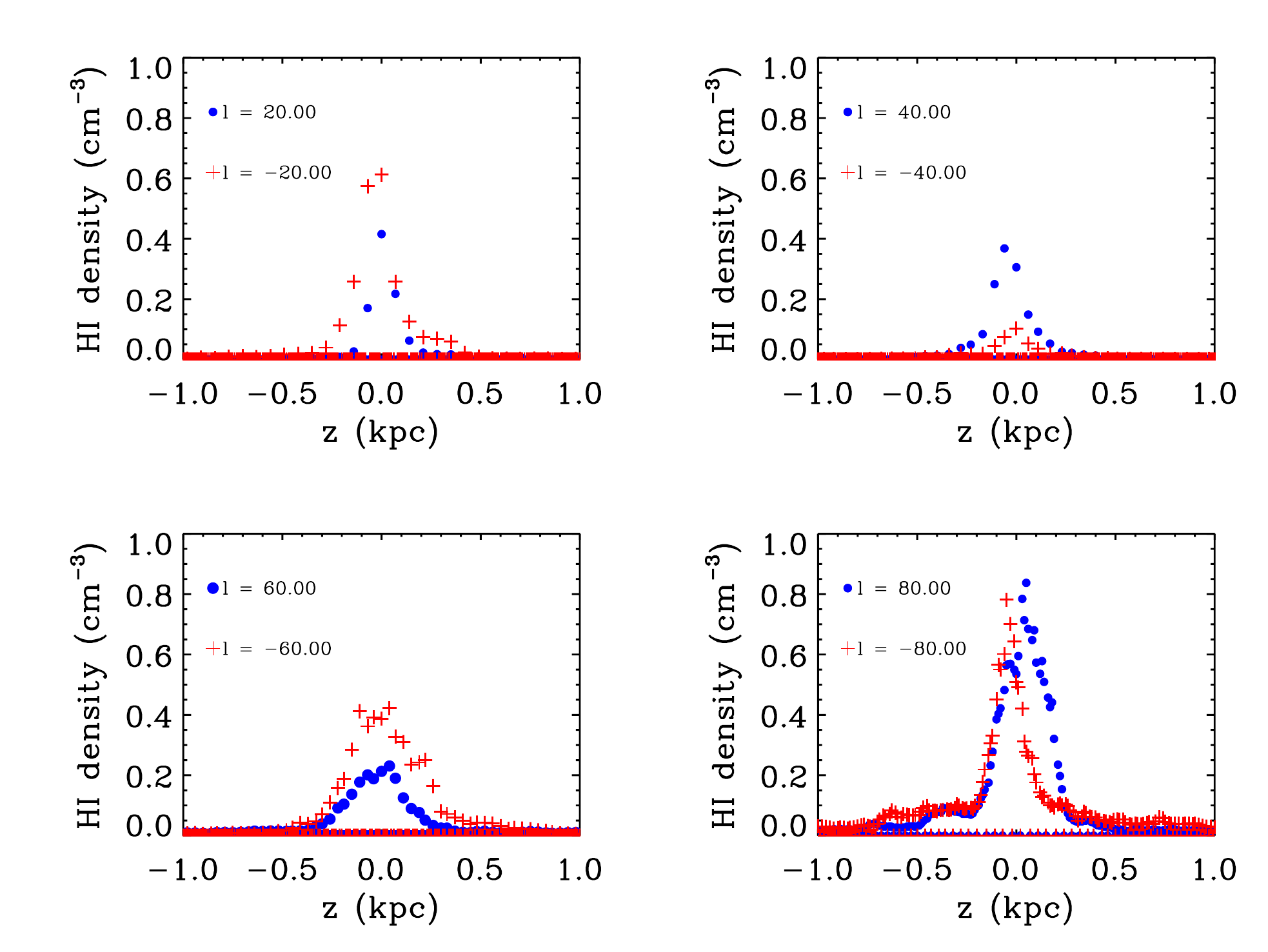}
\end{center}
\caption{Atomic hydrogen number density at tangent points along different directions is plotted versus $z$. The related Galactocentric radius is $R=R_{\odot}|\sin l|$ }
\label{fig.nHI_tangent}
\end{figure}

The hydrogen number density at tangent points along different directions is shown versus height in fig.(\ref{fig.nHI_tangent}). The mid-plane, where the gas density is maximum, almost coincides with the Galactic plane. The vertical distribution of the gas around the mid-plane can be estimated by a Gaussian function of the form $ \exp[{-(\frac{z-z_0}{\sigma_z} })^2]$ where $z_0$ and $\sigma_z$ are, respectively, the mid-plane displacement and the scale height.  The scale height at tangent points varies in a small range between 0.1 to 0.2 kpc. 

\subsection{Inner Galaxy}
\label{sec.inner_galaxy}

Inside the solar orbit, for a given radial velocity there are two heliocentric distances. In order to distribute  the signal among these points, we assume that the vertical extension of the gas at every given $R$ is a Gaussian function whose scale height and mid-plane displacement are obtained from tangent point with the same $R$.  The point whose height is closer to the mid-plane receives more contribution from the signal and has greater number density. The number density at $r_i$, where $i$ indicates either the near or the far point, is obtained by
\be\begin{split}
n_H(r_i,l,b)=&CT_s\ln{\Big(1-\frac{T_b(l,b,v_r)}{T_s}\Big)}\Big|\frac{\delta v_r}{\delta r}\Big| \cr
                      \times & \frac{\exp{[-(\frac{z_i-z_0}{\sigma_z})^2}]}{\Sigma_j \exp{[-(\frac{z_j-z_0}{\sigma_z})^2]}}  
\end{split}\ee


\subsection{Local Gas}

At every direction the radial velocities associated with negative angular velocities or distances far from the Galactic plane ($|z| \gg \sigma_z$) are due to peculiar motions of the local gas. We assume that the gas with peculiar radial velocity is locally distributed by a Gaussian function with radial scale $\sigma_r$ of less than about 4 kpc as follows
\be n_H(r,l,b)= CT_s\!\!\int\! \ln{\Big(1-\frac{T_b(l,b,v_r)}{T_s}\Big)} dv_r \times \frac{e^{-(\frac{r}{\sigma_r})^2}}{\int e^{-(\frac{r}{\sigma_r})^2}dr} \ee
where the integral is performed over peculiar velocities along the related line of sight. The impact of this assumption is marginal, since the amount of gas with peculiar velocity is only $0.033\%$ of the total amount of atomic hydrogen gas in the Galaxy.


\section{Results}
\label{sec.results}

\begin{figure}
\begin{center}
\includegraphics[scale=1.35]{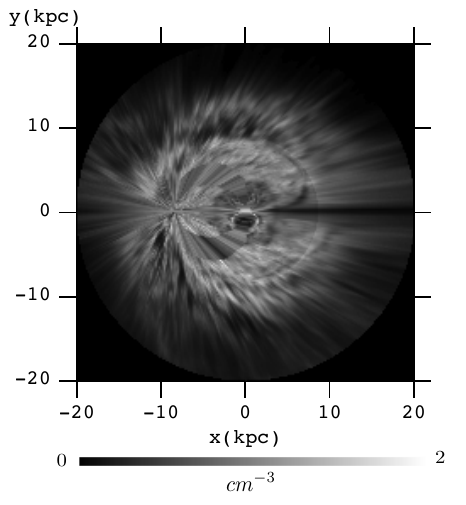}
\end{center}
\caption{Atomic hydrogen density map on the Galactic plane.}
\label{fig.map_z0}
\end{figure}

We follow the method described in section \ref{sec.method} to derive the number density of atomic hydrogen gas as a function of $r$, $l$ and $b$ centered at the Sun. It is transformed into Cartesian coordinates centered at the Galactic center in which the Sun is assumed to be at $(x,y,z) =(-8.5,0,0)$ kpc. The map of number density on the Galactic plane $z=0$ is shown in fig.(\ref{fig.map_z0}). The distribution of gas on the Galactic plane is north-south asymmetric. The density peaks at the Galactic center however  there is a distinct hole right below it. Most of the gas on the Galactic plane is concentrated within the radius of about 10 kpc whereafter it rapidly dilutes away.   

The global properties of the atomic hydrogen gas distribution in the Milky Way are explained in the following sections. 

\subsection{The Warp}

\begin{figure}
\begin{center}
\includegraphics[scale=1.35]{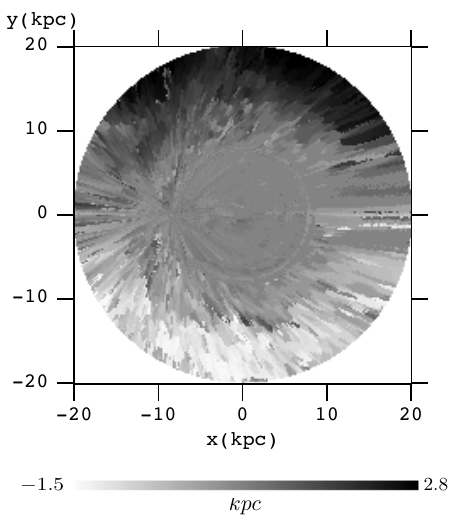}
\end{center}
\caption{The warped Galactic plane. The mid plane displacement has greater positive (negative) values in darker (lighter) regions.}
\label{fig.warp}
\end{figure}

There is a large scale warp in the gas disk of the Milky Way (see \cite{Levine:2006ty} and references therein). 
The map of the mid-plane displacement is shown in fig.(\ref{fig.warp}). Inside solar circle the mid-plane and the Galactic plane coincide pretty well however outside this region the mid-plane is warping. The warp is weak up to radii of about 13 kpc and then it quickly becomes strong. The mid-plane bends up to a height greater than $z_0=2.5$ kpc at R=20 kpc in the north and bends down to $z_0=-1.5$ kpc in the south. 

\begin{figure}
\begin{center}
\includegraphics[scale=0.9]{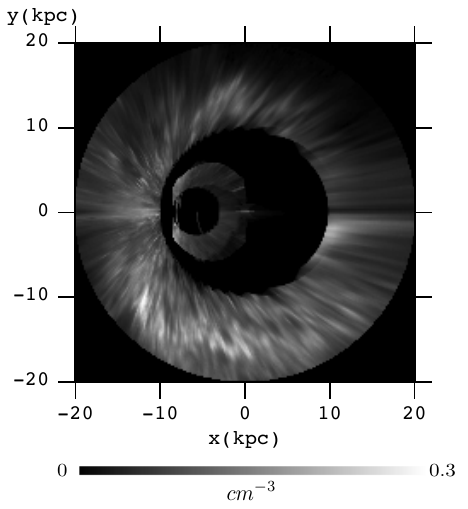}
\hspace{-0.1cm}
\includegraphics[scale=0.9]{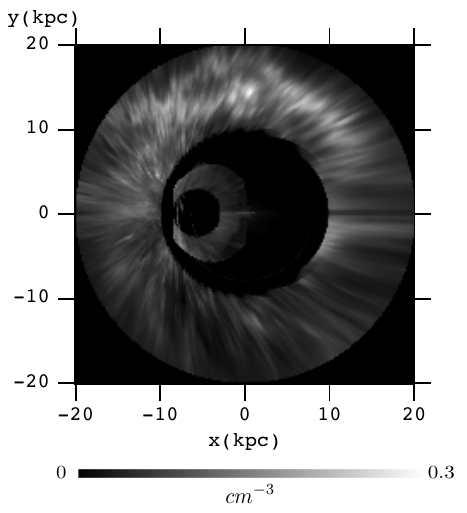} 
\\
\includegraphics[scale=0.9]{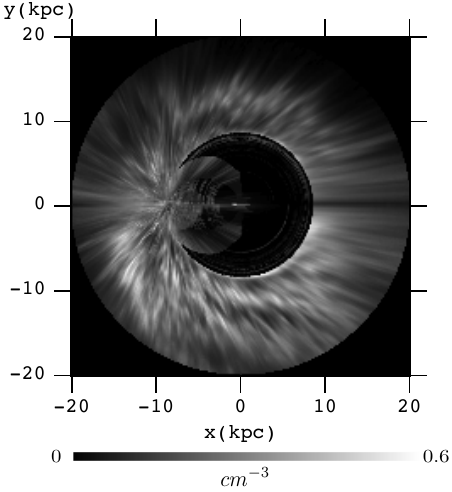}
\hspace{-0.2cm}
\includegraphics[scale=0.9]{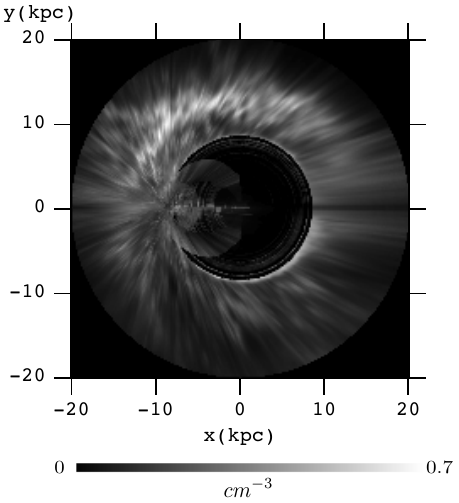}
\end{center}
\caption{Atomic hydrogen density maps at different values of $z$. Plots from left to right are, respectively, due to z=-1,1 kpc on the top panel and due to z=-0.5,0.5 kpc on the bottom panel.}
\label{fig.maps}
\end{figure}

To better illustrate the warping feature of the gas distribution, in fig.\ref{fig.maps} we display the gas number density maps at different heights. 
At negative values of $z$ the dense regions are in the south and at positive values of $z$ they are in the north.

\begin{figure}
\begin{center}
\includegraphics[scale=.4]{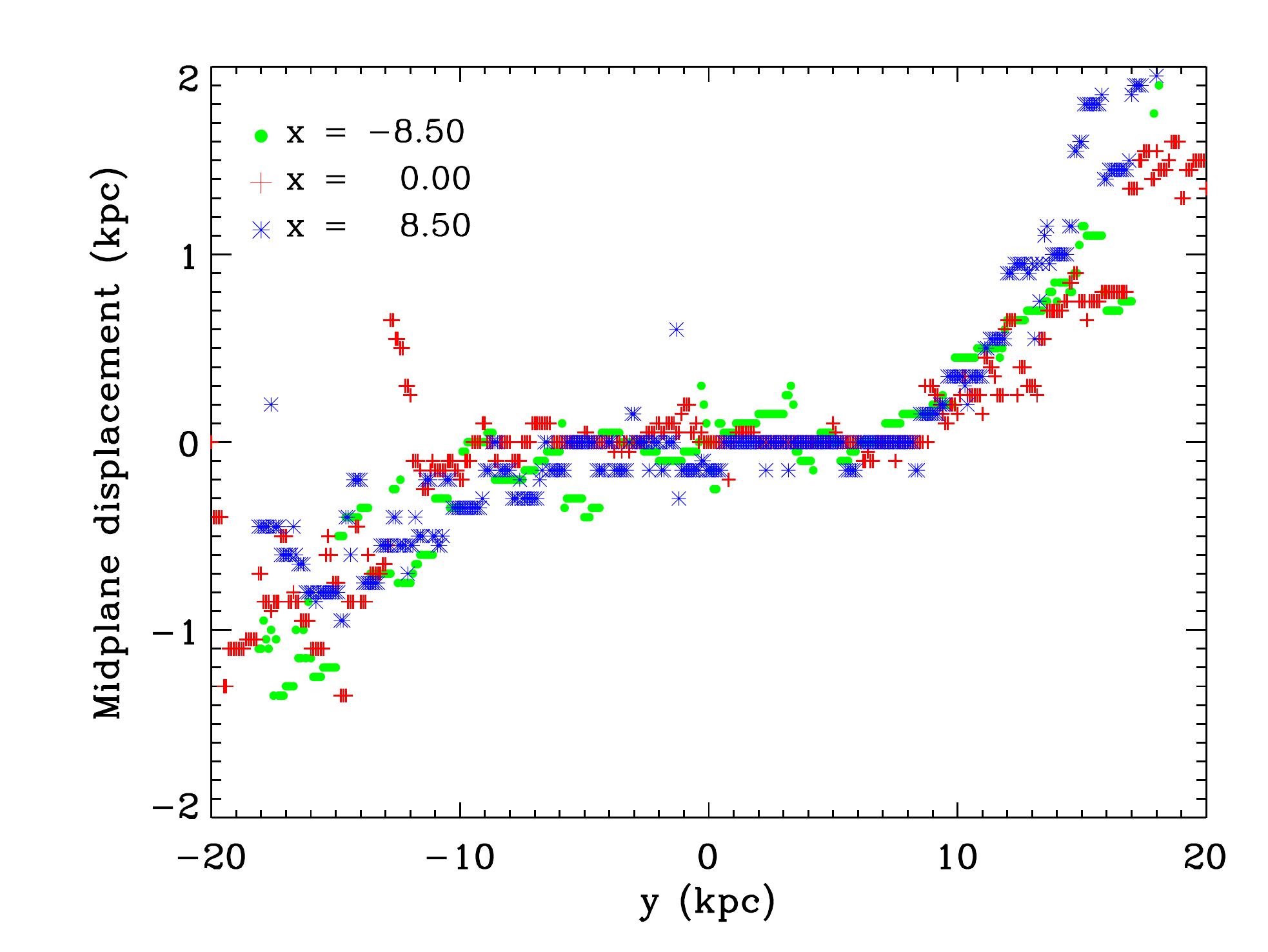}
\end{center}
\caption{The height from the Galactic plane where the number density of the atomic hydrogen gas is maximum.}
\label{fig.midplane_displacement}
\end{figure}

In fig.(\ref{fig.midplane_displacement}) the mid-plane displacement for different values of $x$ is shown against $y$. The general behavior of increasing the mid-plane distance from the Galactic plane outward the Galaxy is asymmetric with more vertical extension in the north.

\subsection{The Flare}

\begin{figure}
\begin{center}
\includegraphics[scale=.4]{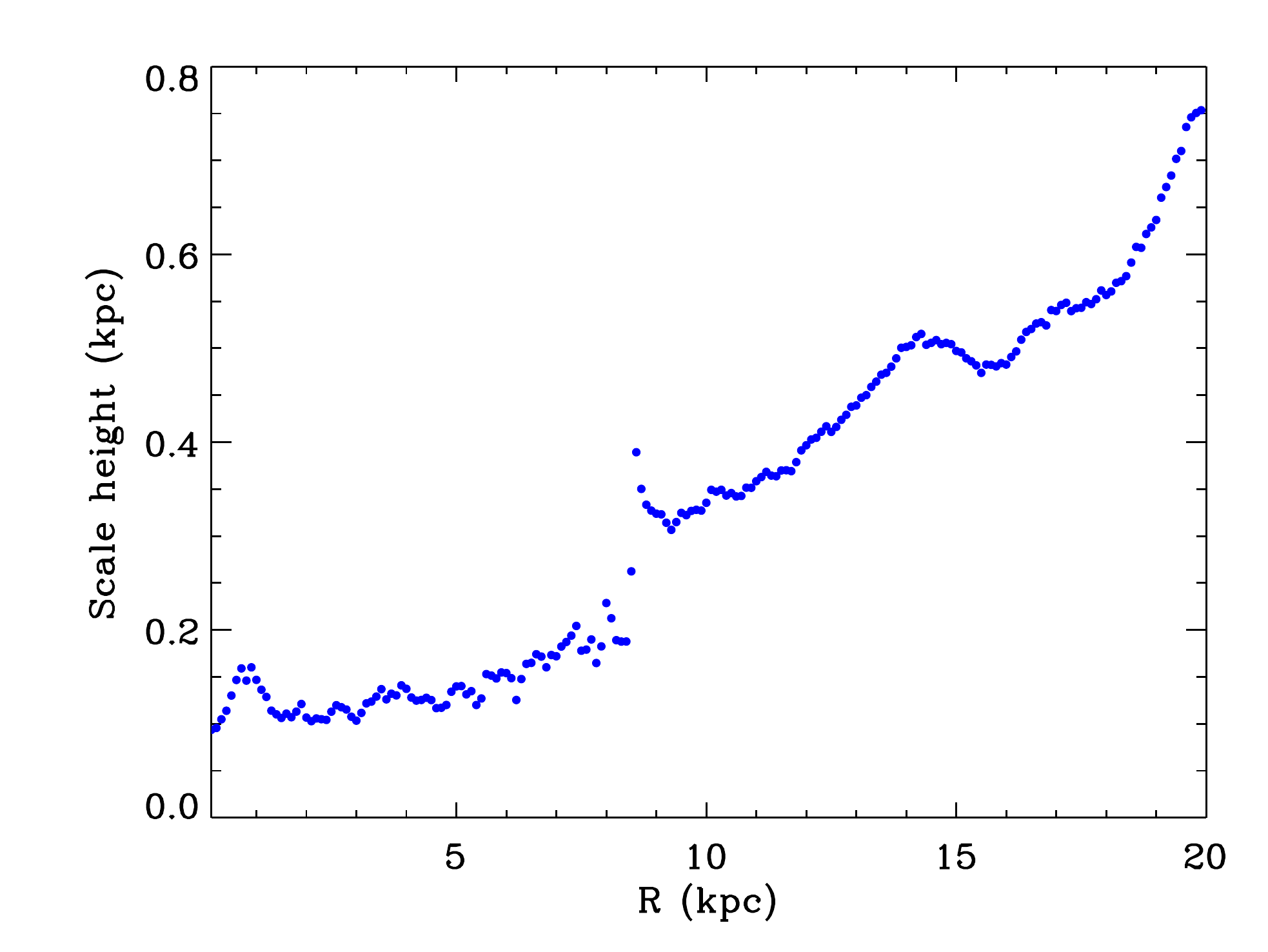}
\hspace{-0.1cm}
\includegraphics[scale=.4]{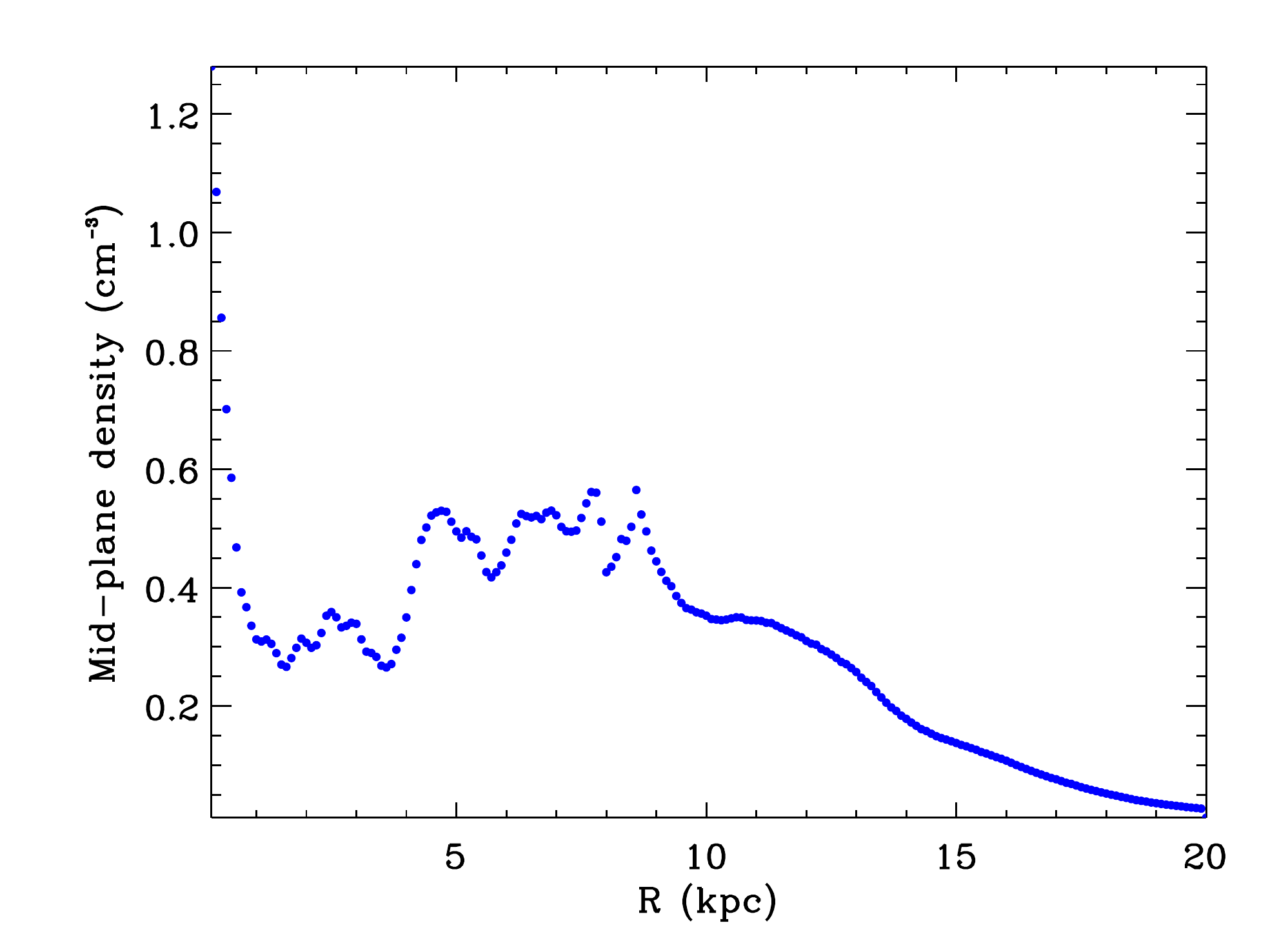}
\end{center}
\caption{Top: The average scale height versus Galactocentric radius. Bottom: The mid-plane density versus $R$.}
\label{fig.scale_height}
\end{figure}

The balance of gravitational force against the pressure force determines the thickness of the atomic hydrogen disk. The average scale height, which is defined as the distance over which the number density decreases by a factor of $e$,  shows a clear flaring \cite{Kalberla:2007sr, 1992AJ....103.1552M, 1990A&A...230...21W, 1979A&A....76...24C}. It increases from about 0.2 kpc in the inner Galaxy up to 0.75 kpc at R=20 kpc as shown in fig.(\ref{fig.scale_height}, top panel).  We also show the mid-plane density in fig.(\ref{fig.scale_height}, bottom panel). It peaks at the Galactic center and has fluctuations in the inner part, then falls down in the outer Galaxy. 

\subsection{Spiral Structure}

\begin{figure}
\begin{center}
\includegraphics[scale=1.35]{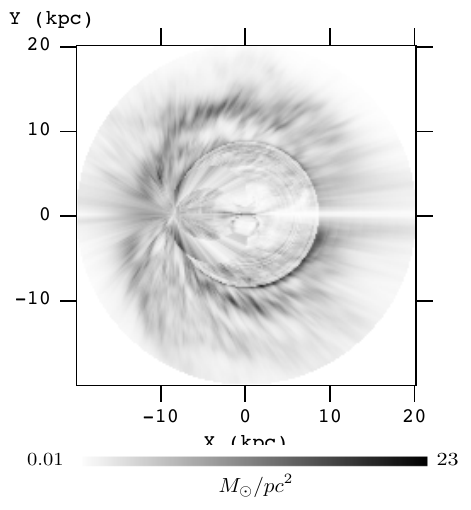}
\end{center}
\caption{The surface density map. The  spiral arms can be traced as regions with over densities.}
\label{fig.spiral}
\end{figure}

The spiral structure can be traced in the surface density distribution as regions with over densities \cite{Levine:2006yv, 2004ApJ...607L.127M}. In fig.(\ref{fig.spiral}) we show the surface density map,
\be
\Sigma(x,y)=\int dz n_H(x,y,z)
\ee
in which several spiral arms are evident. There is one large spiral arm in the north, the so-called Outer arm. In the southern half, the so-called Sagittarius-Carina arm close to the solar circle is prominent. The so-called Perseus arm in the south extends to the north and connects to the Outer arm. 

\section{Conclusions}
\label{sec.summary}

In this paper we present  a new model for three dimensional distribution of atomic hydrogen gas in the Milky Way. The most recent data on 21cm line emission provided by the LAB survey is used. To convert the observed brightness temperature distribution to the volume density distribution, we assume a purely circular rotation curve. We expect it to be a reasonable approximation, except in the central region of the Galaxy where the central bar exists. The vertical distribution of gas around the Galactic plane is estimated as a Gaussian function. 

The overall structure of the gas distribution discloses the warping of the Galactic plane outside the solar circle. The bending becomes strong at Galactocentric radii greater than about 13 kpc.  The thickness of the gas disk flares outward the Galaxy. At the same time, the mid-plane density falls down in the outer part of the Galaxy. Several spiral arms can be traced in the surface density map. It is found that the total mass within a radius of 20 kpc is $4.3 \times 10^9 M_{\odot}$ and only $0.033 \%$ of that is due to local gas with peculiar velocities. 

The derived distribution of atomic hydrogen gas can be used to study the propagation of cosmic rays within the Galaxy. It can also be used to evaluate the diffuse gamma ray maps. Indeed, the structures of the gas distribution can be identified by high angular resolution maps of the {\it Fermi} gamma ray telescope.   

This model is publicly available at  the following link {\tt http://people.sissa.it/$\sim$tavakoli }


\acknowledgments

The author is  grateful to Piero Ullio, Ilias Cholis and Carmelo Evoli for valuable discussions and comments.


\bibliography{HI}

\begin{thebibliography}{26}
\expandafter\ifx\csname natexlab\endcsname\relax\def\natexlab#1{#1}\fi
\expandafter\ifx\csname bibnamefont\endcsname\relax
  \def\bibnamefont#1{#1}\fi
\expandafter\ifx\csname bibfnamefont\endcsname\relax
  \def\bibfnamefont#1{#1}\fi
\expandafter\ifx\csname citenamefont\endcsname\relax
  \def\citenamefont#1{#1}\fi
\expandafter\ifx\csname url\endcsname\relax
  \def\url#1{\texttt{#1}}\fi
\expandafter\ifx\csname urlprefix\endcsname\relax\def\urlprefix{URL }\fi
\providecommand{\bibinfo}[2]{#2}
\providecommand{\eprint}[2][]{\url{#2}}

\bibitem[{\citenamefont{{Kalberla}}(2003)}]{2003ApJ...588..805K}
\bibinfo{author}{\bibfnamefont{P.~M.~W.} \bibnamefont{{Kalberla}}},
  \bibinfo{journal}{\apj} \textbf{\bibinfo{volume}{588}}, \bibinfo{pages}{805}
  (\bibinfo{year}{2003}).

\bibitem[{\citenamefont{{Kalberla} and {Kerp}}(2009)}]{2009ARA&A..47...27K}
\bibinfo{author}{\bibfnamefont{P.~M.~W.} \bibnamefont{{Kalberla}}}
  \bibnamefont{and} \bibinfo{author}{\bibfnamefont{J.}~\bibnamefont{{Kerp}}},
  \bibinfo{journal}{\araa} \textbf{\bibinfo{volume}{47}}, \bibinfo{pages}{27}
  (\bibinfo{year}{2009}).

\bibitem[{\citenamefont{Nakanishi and Sofue}(2003)}]{Nakanishi:2003eb}
\bibinfo{author}{\bibfnamefont{H.}~\bibnamefont{Nakanishi}} \bibnamefont{and}
  \bibinfo{author}{\bibfnamefont{Y.}~\bibnamefont{Sofue}},
  \bibinfo{journal}{Publ.Astron.Soc.Jap.} \textbf{\bibinfo{volume}{55}},
  \bibinfo{pages}{191} (\bibinfo{year}{2003}), \eprint{astro-ph/0304338}.

\bibitem[{\citenamefont{{Dickey} and {Lockman}}(1990)}]{1990ARA&A..28..215D}
\bibinfo{author}{\bibfnamefont{J.~M.} \bibnamefont{{Dickey}}} \bibnamefont{and}
  \bibinfo{author}{\bibfnamefont{F.~J.} \bibnamefont{{Lockman}}},
  \bibinfo{journal}{\araa} \textbf{\bibinfo{volume}{28}}, \bibinfo{pages}{215}
  (\bibinfo{year}{1990}).

\bibitem[{\citenamefont{{Gordon} and {Burton}}(1976)}]{1976ApJ...208..346G}
\bibinfo{author}{\bibfnamefont{M.~A.} \bibnamefont{{Gordon}}} \bibnamefont{and}
  \bibinfo{author}{\bibfnamefont{W.~B.} \bibnamefont{{Burton}}},
  \bibinfo{journal}{\apj} \textbf{\bibinfo{volume}{208}}, \bibinfo{pages}{346}
  (\bibinfo{year}{1976}).

\bibitem[{\citenamefont{Kalberla et~al.}(2005)\citenamefont{Kalberla, Burton,
  Hartmann, Arnal, Bajaja et~al.}}]{Kalberla:2005ts}
\bibinfo{author}{\bibfnamefont{P.~M.} \bibnamefont{Kalberla}},
  \bibinfo{author}{\bibfnamefont{W.}~\bibnamefont{Burton}},
  \bibinfo{author}{\bibfnamefont{D.}~\bibnamefont{Hartmann}},
  \bibinfo{author}{\bibfnamefont{E.}~\bibnamefont{Arnal}},
  \bibinfo{author}{\bibfnamefont{E.}~\bibnamefont{Bajaja}},
  \bibnamefont{et~al.}, \bibinfo{journal}{Astron.Astrophys.}
  \textbf{\bibinfo{volume}{440}}, \bibinfo{pages}{775} (\bibinfo{year}{2005}),
  \eprint{astro-ph/0504140}.

\bibitem[{\citenamefont{{Arnal} et~al.}(2000)\citenamefont{{Arnal}, {Bajaja},
  {Larrarte}, {Morras}, and {P{\"o}ppel}}}]{2000A&AS..142...35A}
\bibinfo{author}{\bibfnamefont{E.~M.} \bibnamefont{{Arnal}}},
  \bibinfo{author}{\bibfnamefont{E.}~\bibnamefont{{Bajaja}}},
  \bibinfo{author}{\bibfnamefont{J.~J.} \bibnamefont{{Larrarte}}},
  \bibinfo{author}{\bibfnamefont{R.}~\bibnamefont{{Morras}}}, \bibnamefont{and}
  \bibinfo{author}{\bibfnamefont{W.~G.~L.} \bibnamefont{{P{\"o}ppel}}},
  \bibinfo{journal}{\aaps} \textbf{\bibinfo{volume}{142}}, \bibinfo{pages}{35}
  (\bibinfo{year}{2000}).

\bibitem[{\citenamefont{Bajaja et~al.}(2005)\citenamefont{Bajaja, Arnal,
  Larrarte, Morras, Poppel et~al.}}]{Bajaja:2005tn}
\bibinfo{author}{\bibfnamefont{E.}~\bibnamefont{Bajaja}},
  \bibinfo{author}{\bibfnamefont{E.}~\bibnamefont{Arnal}},
  \bibinfo{author}{\bibfnamefont{J.}~\bibnamefont{Larrarte}},
  \bibinfo{author}{\bibfnamefont{R.}~\bibnamefont{Morras}},
  \bibinfo{author}{\bibfnamefont{W.}~\bibnamefont{Poppel}},
  \bibnamefont{et~al.}, \bibinfo{journal}{Astron.Astrophys.}
  (\bibinfo{year}{2005}), \eprint{astro-ph/0504136}.

\bibitem[{\citenamefont{Hartmann and Burton}(1997)}]{Hartmann:1997}
\bibinfo{author}{\bibfnamefont{D.}~\bibnamefont{Hartmann}} \bibnamefont{and}
  \bibinfo{author}{\bibfnamefont{W.}~\bibnamefont{Burton}},
  \bibinfo{journal}{Cambridge, UK: Cambridge Univ. Press}
  (\bibinfo{year}{1997}).

\bibitem[{\citenamefont{Longair}(2002)}]{Longair}
\bibinfo{author}{\bibfnamefont{M.~S.} \bibnamefont{Longair}},
  \emph{\bibinfo{title}{High Energy Astrophysics}}
  (\bibinfo{publisher}{Cambridge University Press}, \bibinfo{year}{2002}).

\bibitem[{\citenamefont{Binney and Merrifield}(1998)}]{Binney}
\bibinfo{author}{\bibfnamefont{J.}~\bibnamefont{Binney}} \bibnamefont{and}
  \bibinfo{author}{\bibfnamefont{M.}~\bibnamefont{Merrifield}},
  \emph{\bibinfo{title}{Galactic Astronomy}} (\bibinfo{publisher}{Princeton
  University Press}, \bibinfo{year}{1998}).

\bibitem[{\citenamefont{Mihalas and Binney}(1981)}]{Mihalas1981}
\bibinfo{author}{\bibfnamefont{D.}~\bibnamefont{Mihalas}} \bibnamefont{and}
  \bibinfo{author}{\bibfnamefont{J.}~\bibnamefont{Binney}},
  \emph{\bibinfo{title}{Galactic Astronomy, Structure and Kinematics}}
  (\bibinfo{publisher}{Freeman}, \bibinfo{year}{1981}).

\bibitem[{\citenamefont{Choudhuri}(2010)}]{Choudhuri}
\bibinfo{author}{\bibfnamefont{A.~R.} \bibnamefont{Choudhuri}},
  \emph{\bibinfo{title}{Astrophysics for Physicists}}
  (\bibinfo{publisher}{Cambridge University Press}, \bibinfo{year}{2010}).

\bibitem[{\citenamefont{Strasser and Taylor}(2004)}]{Strasser:2004uh}
\bibinfo{author}{\bibfnamefont{S.~T.} \bibnamefont{Strasser}} \bibnamefont{and}
  \bibinfo{author}{\bibfnamefont{A.}~\bibnamefont{Taylor}},
  \bibinfo{journal}{Astrophys.J.} \textbf{\bibinfo{volume}{603}},
  \bibinfo{pages}{560} (\bibinfo{year}{2004}), \eprint{astro-ph/0401248}.

\bibitem[{\citenamefont{Dickey et~al.}(2009)\citenamefont{Dickey, Strasser,
  Gaensler, Haverkorn, Kavars et~al.}}]{Dickey:2009mu}
\bibinfo{author}{\bibfnamefont{J.~M.} \bibnamefont{Dickey}},
  \bibinfo{author}{\bibfnamefont{S.}~\bibnamefont{Strasser}},
  \bibinfo{author}{\bibfnamefont{B.}~\bibnamefont{Gaensler}},
  \bibinfo{author}{\bibfnamefont{M.}~\bibnamefont{Haverkorn}},
  \bibinfo{author}{\bibfnamefont{D.}~\bibnamefont{Kavars}},
  \bibnamefont{et~al.}, \bibinfo{journal}{Astrophys.J.}
  \textbf{\bibinfo{volume}{693}}, \bibinfo{pages}{1250} (\bibinfo{year}{2009}),
  \eprint{0901.0968}.

\bibitem[{\citenamefont{Johannesson et~al.}(2010)\citenamefont{Johannesson,
  Moskalenko, and Digel}}]{Johannesson:2010fr}
\bibinfo{author}{\bibfnamefont{G.}~\bibnamefont{Johannesson}},
  \bibinfo{author}{\bibfnamefont{I.}~\bibnamefont{Moskalenko}},
  \bibnamefont{and} \bibinfo{author}{\bibfnamefont{S.}~\bibnamefont{Digel}}
  (\bibinfo{collaboration}{Fermi LAT Collaboration}) (\bibinfo{year}{2010}),
  \eprint{1002.0081}.

\bibitem[{\citenamefont{Kalberla et~al.}(2007)\citenamefont{Kalberla, Dedes,
  Kerp, and Haud}}]{Kalberla:2007sr}
\bibinfo{author}{\bibfnamefont{P.}~\bibnamefont{Kalberla}},
  \bibinfo{author}{\bibfnamefont{L.}~\bibnamefont{Dedes}},
  \bibinfo{author}{\bibfnamefont{J.}~\bibnamefont{Kerp}}, \bibnamefont{and}
  \bibinfo{author}{\bibfnamefont{U.}~\bibnamefont{Haud}}
  (\bibinfo{year}{2007}), \eprint{0704.3925}.

\bibitem[{\citenamefont{Clemens}(1985)}]{Clemens:1985}
\bibinfo{author}{\bibfnamefont{D.}~\bibnamefont{Clemens}},
  \bibinfo{journal}{Astrophys.J.} \textbf{\bibinfo{volume}{295}},
  \bibinfo{pages}{422} (\bibinfo{year}{1985}).

\bibitem[{\citenamefont{Levine et~al.}(2006{\natexlab{a}})\citenamefont{Levine,
  Blitz, and Heiles}}]{Levine:2006ty}
\bibinfo{author}{\bibfnamefont{E.~S.} \bibnamefont{Levine}},
  \bibinfo{author}{\bibfnamefont{L.}~\bibnamefont{Blitz}}, \bibnamefont{and}
  \bibinfo{author}{\bibfnamefont{C.}~\bibnamefont{Heiles}},
  \bibinfo{journal}{Astrophys.J.} \textbf{\bibinfo{volume}{643}},
  \bibinfo{pages}{881} (\bibinfo{year}{2006}{\natexlab{a}}),
  \eprint{astro-ph/0601697}.

\bibitem[{\citenamefont{McClure-Griffiths and
  Dickey}(2007)}]{McClureGriffiths:2007ts}
\bibinfo{author}{\bibfnamefont{N.}~\bibnamefont{McClure-Griffiths}}
  \bibnamefont{and} \bibinfo{author}{\bibfnamefont{J.~M.}
  \bibnamefont{Dickey}}, \bibinfo{journal}{Astrophys.J.}
  \textbf{\bibinfo{volume}{671}}, \bibinfo{pages}{427} (\bibinfo{year}{2007}),
  \eprint{0708.0870}.

\bibitem[{\citenamefont{Malhotra}(1995)}]{Malhotra:1994qj}
\bibinfo{author}{\bibfnamefont{S.}~\bibnamefont{Malhotra}},
  \bibinfo{journal}{Astrophys.J.} \textbf{\bibinfo{volume}{448}},
  \bibinfo{pages}{138} (\bibinfo{year}{1995}), \eprint{astro-ph/9411088}.

\bibitem[{\citenamefont{{Merrifield}}(1992)}]{1992AJ....103.1552M}
\bibinfo{author}{\bibfnamefont{M.~R.} \bibnamefont{{Merrifield}}},
  \bibinfo{journal}{\aj} \textbf{\bibinfo{volume}{103}}, \bibinfo{pages}{1552}
  (\bibinfo{year}{1992}).

\bibitem[{\citenamefont{Wouterloot et~al.}(1990)\citenamefont{Wouterloot,
  Brand, Burton, and Kwee}}]{1990A&A...230...21W}
\bibinfo{author}{\bibfnamefont{J.~G.~A.} \bibnamefont{Wouterloot}},
  \bibinfo{author}{\bibfnamefont{J.}~\bibnamefont{Brand}},
  \bibinfo{author}{\bibfnamefont{W.~B.} \bibnamefont{Burton}},
  \bibnamefont{and} \bibinfo{author}{\bibfnamefont{K.~K.} \bibnamefont{Kwee}},
  \bibinfo{journal}{\aap} \textbf{\bibinfo{volume}{230}}, \bibinfo{pages}{21}
  (\bibinfo{year}{1990}).

\bibitem[{\citenamefont{{Celnik} et~al.}(1979)\citenamefont{{Celnik}, {Rohlfs},
  and {Braunsfurth}}}]{1979A&A....76...24C}
\bibinfo{author}{\bibfnamefont{W.}~\bibnamefont{{Celnik}}},
  \bibinfo{author}{\bibfnamefont{K.}~\bibnamefont{{Rohlfs}}}, \bibnamefont{and}
  \bibinfo{author}{\bibfnamefont{E.}~\bibnamefont{{Braunsfurth}}},
  \bibinfo{journal}{\aap} \textbf{\bibinfo{volume}{76}}, \bibinfo{pages}{24}
  (\bibinfo{year}{1979}).

\bibitem[{\citenamefont{Levine et~al.}(2006{\natexlab{b}})\citenamefont{Levine,
  Blitz, and Heiles}}]{Levine:2006yv}
\bibinfo{author}{\bibfnamefont{E.~S.} \bibnamefont{Levine}},
  \bibinfo{author}{\bibfnamefont{L.}~\bibnamefont{Blitz}}, \bibnamefont{and}
  \bibinfo{author}{\bibfnamefont{C.}~\bibnamefont{Heiles}},
  \bibinfo{journal}{Science}  (\bibinfo{year}{2006}{\natexlab{b}}),
  \eprint{astro-ph/0605728}.

\bibitem[{\citenamefont{{McClure-Griffiths}
  et~al.}(2004)\citenamefont{{McClure-Griffiths}, {Dickey}, {Gaensler}, and
  {Green}}}]{2004ApJ...607L.127M}
\bibinfo{author}{\bibfnamefont{N.~M.} \bibnamefont{{McClure-Griffiths}}},
  \bibinfo{author}{\bibfnamefont{J.~M.} \bibnamefont{{Dickey}}},
  \bibinfo{author}{\bibfnamefont{B.~M.} \bibnamefont{{Gaensler}}},
  \bibnamefont{and} \bibinfo{author}{\bibfnamefont{A.~J.}
  \bibnamefont{{Green}}}, \bibinfo{journal}{\apjl}
  \textbf{\bibinfo{volume}{607}}, \bibinfo{pages}{L127} (\bibinfo{year}{2004}),
  \eprint{arXiv:astro-ph/0404448}.

\end{thebibliography}

\end{document}